 \documentclass[prd,preprintnumbers,amsmath,amssymb,floatfix]{revtex4}

\usepackage{graphicx}
\usepackage{dcolumn}
\usepackage{bm}
\usepackage{amssymb}
\usepackage{epsfig}
\usepackage{color}

\newcommand{\ba}{\begin{eqnarray}}
\newcommand{\ea}{\end{eqnarray}}
\newcommand{\be}{\begin{equation}}
\newcommand{\ee}{\end{equation}}
\newcommand{\bdisplay}{\begin{displaymath}}
\newcommand{\edisplay}{\end{displaymath}}

\newcommand{\eq}[1]{Eq.\,(\ref{#1})}

\newcommand{\pbar}{\bar p}

\newcommand{\sigin}{\sigma_{\rm inel}}
\newcommand{\sigel}{\sigma_{\rm el}}
\newcommand{\sigtot}{\sigma_{\rm tot}}
\begin{document}

\title{Experimental Confirmation that the Proton is Asymptotically a Black Disk}  
\author{Martin~M.~Block}
\affiliation{Department of Physics and Astronomy, Northwestern University, 
Evanston, IL 60208}
\author{Francis Halzen}
\affiliation{Department of Physics, University of Wisconsin, Madison, WI 53706}
\date{\today}

\begin{abstract}
Although experimentally accessible energies can not probe `asymptopia', recent measurements of` inelastic $pp$ cross sections at the LHC at 7000 GeV and by Auger at 57000 GeV allow us to conclude that: i) both $\sigin$ and $\sigtot$, the inelastic and total cross sections for $pp$ and $\bar p p$ interactions, saturate the Froissart bound of  $\ln^2 s$, ii) when  $s\rightarrow \infty$, the ratio $\sigin/\sigtot$ is experimentally determined to be $0.509\pm 0.021$,  consistent with the value 0.5 required by black disk at infinite energies, and iii) when $s\rightarrow \infty$, the forward scattering amplitude becomes purely imaginary, another requirement for the proton to become a totally absorbing black disk. Experimental verification of the hypotheses of analyticity and unitarity over the center of mass energy range $6\le \sqrt s\le 57000$ GeV are discussed. In QCD, the black disk is naturally made of gluons;  our results  suggest that the lowest-lying glueball mass is $2.97\pm 0.03$ GeV.
 \end{abstract}

\maketitle


{\em Introduction:} We discuss the implication of three new measurements of the high energy $pp$ inelastic cross sections, $\sigin(\sqrt s)$, where $\sqrt s$ is the cms (center of mass) energy.  At $\sqrt s= 7000$ GeV, the Atlas collaboration \cite{Atlas} reports $\sigin =69.4\pm 2.4 \ {(\rm expt.})\pm 6.9 \ {(\rm extr.)}$ mb, with (expt.) and (extr.)  the total experimental and extrapolation errors.  The CMS collaboration \cite{CMS}, using a completely different technique,  measures $\sigin =68.0\pm 2.0\ {(\rm syst.)}\pm 2.4 \pm \ {(\rm lum.)}\pm {(\rm extr.)}$, where (syst.) is the systematic error, (lum.) the error in luminosity and (extr.) is the extrapolation error for missing single and double diffraction events. Most recently, the Pierre Auger Observatory collaboration \cite {Auger} reported a measurement of $\sigma_{\rm inel}^{\rm p-air}$, the inelastic p-air cross section at $\sqrt s=57000\pm 6000$ GeV. This measurement, after  correction for a 25\% ${\rm He^4}$ contamination in a cosmic ray beam consisting mostly of protons at that energy, was converted by a Glauber calculation into the $pp$ inelastic cross section \cite{Auger}, $\sigin=90\pm7 \ {(\rm stat.)}\pm\ \! ^9_{11} \ {(\rm syst.)}\pm 1.5\ {(\rm Glaub.)}$, with (stat.) the statistical, (syst.) the systematic errors and  (Glaub.)  the estimated error in the Glauber calculation. With a cosmic ray measurement at 57000 GeV  it is likely that  we are now experimentally as close to asymptopia (defined here as the energy behavior of hadron-proton cross sections near $s\rightarrow\infty$)  as we will ever get. 

Block and Halzen (BH) \cite{BH,physicsreports} have made an analyticity constrained amplitude fit to lower energy data ($6\le \sqrt s
\le 2000$ GeV) that shows that $\sigtot$ for $\bar pp$ and $pp$ asymptotically saturates the Froissart bound \cite{froissart}. This  note  exploits the new  higher energy measurements of $\sigin$  in order to make accurate  predictions at asymptotia based only on measurements of $pp$ and $\bar p p$ cross sections  in the energy range $6\le \sqrt s \le 57000$ GeV. While the analyticity constrained amplitude model of BH \cite{BH,physicsreports} yields the total cross sections and the $\rho$-value, the ratio of real and the imaginary parts of the forward scattering amplitude, an eikonal model, dubbed the `Aspen' model \cite{aspenmodel}, will be used to obtain the ratio of the inelastic to total cross sections,  $r(\sqrt s)\equiv \sigin(\sqrt s) /\sigtot(\sqrt s)$. We will show that the resulting $\rho$-value and  the ratio of $\sigin /\sigtot$ at $\sqrt s =\infty$ are consistent with the proton being an expanding black disk, presumably of gluons; our fits to $\sigin$ and $\sigel$ will allow us to infer  a lowest-lying glueball mass of $2.97\pm 0.03$ GeV. Furthermore, we will show that both  the Martin-Froissart bound \cite{martin1,froissart} on the $pp$ and $\bar p p$ total  cross sections and the Martin bound \cite{martin2} on the  $pp$ and $\bar pp$ inelastic cross sections are saturated, from $6\le\sqrt s\le 57000$ GeV.

\begin{figure}[h,t,b] 
\begin{center}
\mbox{\epsfig{file=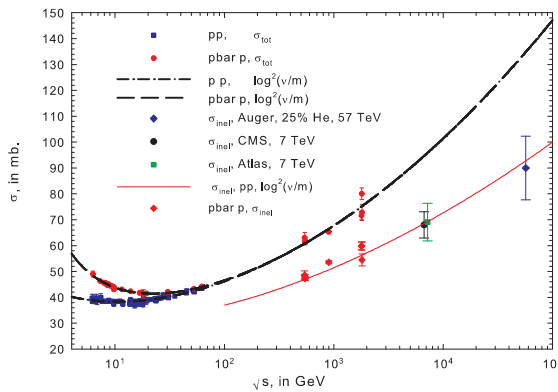
,width=3in%
,bbllx=150pt,bblly=375pt,bburx=595pt,bbury=680pt,clip=%
}}
\end{center}
\caption[]{
The fitted total cross section, $\sigtot$,  for $\bar pp$ (dashed curve)  and $pp$ (dot-dashed curve)  from \eq{sigmapmpp}, in mb vs. $\sqrt s$, the cms energy in GeV, taken from BH  \cite{BH}. The $\bar pp$ data used in the fit  are the (red) circles and the $pp$ data are the (blue) squares. The fitted data were anchored by values of $\sigtot^{\bar pp}$ and $\sigtot^{pp}$, together with the energy derivatives  ${d\sigtot^{\bar pp}/ d\nu}$ and ${d\sigtot^{pp}/ d\nu}$ at 6 GeV using FESR, as described in Ref. \cite{BH}.  The lowest  (red) solid curve that starts at 100 GeV is our {\em predicted} inelastic cross section from \eq{finalinelastic}, $\sigin$, in mb,  vs. $\sqrt s$, in GeV. The lowest energy inelastic data, the  $\bar pp$ (red) diamonds, were {\em not} used in the fit, nor were the 3 high energy $pp$ inelastic measurements, the (black) circle  CMS value, the (green) square Atlas measurement and the (blue) diamond Auger measurement. As clearly seen, our inelastic prediction from \eq{finalinelastic}, which also asymptotically behaves as $\ln^2(s)$, is in excellent agreement with the new measurements of the inelastic cross section at very high energy.
\label{fig:pp}
}
\end{figure}

\begin{figure}[h,t,b] 
\begin{center}
\mbox{\epsfig{file=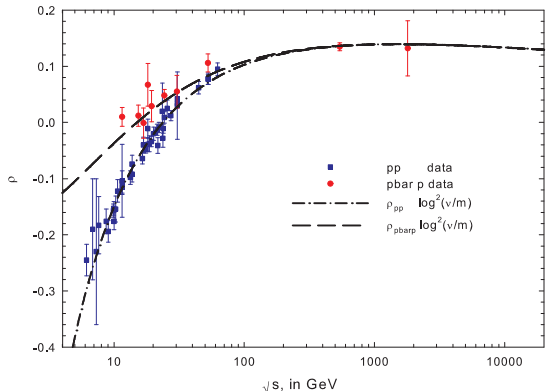
,width=3in%
,bbllx=100pt,bblly=300pt,bburx=530pt,bbury=600pt,clip=%
}}
\end{center}
\caption[]{
 The fitted $\rho$-value,  for $\bar pp$ (dashed curve)  and $pp$ (dot-dashed curve)  from \eq{sigmapmpp} vs. $\sqrt s$, the cms energy in GeV. The $\bar pp$ data used in the fit  are the (red) circles and the $pp$ data are the (blue) squares.\label{fig:rho}
}
\end{figure}

{\em The Analytic Amplitude Model:}
Using this approach, BH was able to claim accurate predictions of  the forward $pp$ ($\bar pp$)  scattering properties,
$
\sigma_{\rm tot}\equiv{4\pi\over p} {\rm Im}f(\theta_L=0)$ and 
$\rho\equiv{{\rm Re}f(\theta_L=0 )\over {\rm Im}f(\theta_L=0)},
$
using the analyticity-constrained analytic amplitude model\cite{BH} that saturates the Froissart bound \cite{froissart};  here $f(\theta_L)$ is the $pp$  laboratory scattering amplitude with $\theta_L$, the laboratory scattering angle and $p$ is the laboratory momentum. By saturation of the Froissart bound, we mean  that the total cross section $\sigtot$ rises as $\ln^2 s$. Furthermore the use of analyticity constraints allows one to anchor  fits at 6 GeV to the  very accurate  low energy cross section measurements  between 4 and 6 GeV in the spirit of Finite Energy Sum Rules (FESR)\cite {FESR}. A local fit is made of the experimental values of $\sigma^{\pm}$ between 4 and 6 GeV, for both $\bar pp$ and $pp$, from which BH  \cite{BH} derive precise 6 GeV `anchor-points' for $\sigma^{\pm}$ and their energy derivatives in \eq{sigmapmpp}. The results are actually consistent with those obtained with old-fashioned FESR\cite{igi}. The model parameterizes the even and odd (under crossing) cross sections and fits \cite{BH} 4 experimental quantities, $\sigma_{\bar pp}(\nu), \sigma_{p p}(\nu), \rho_{\bar pp}(\nu)$ and $\rho_{p p}(\nu)$ to the high energy parameterizations
\ba
\sigma^\pm(\nu)&=&\sigma^0(\nu)\pm\  \delta\left({\nu\over m}\right)^{\alpha -1},\label{sigmapmpp}\\
\rho^\pm(\nu)&=&{1\over\sigma^\pm(\nu)}\left\{\frac{\pi}{2}c_1+c_2\pi \ln\left(\frac{\nu}{m}\right)-\beta_{\cal P'}\cot({\pi\mu\over 2})\left(\frac{\nu}{m}\right)^{\mu -1}+\frac{4\pi}{\nu}f_+(0)\right.
\left.\pm \delta\tan({\pi\alpha\over 2})\left({\nu\over m}\right)^{\alpha -1} \right\}\label{rhopmpp},
\ea
where the upper sign is for $pp$ and the lower sign is for  $\bar pp$,  and, for high energies, ${\nu / m}\simeq{s/ 2m^2}$. Here the even amplitude cross section $\sigma^0$ is given by
\ba
\sigma^0(\nu)&\equiv&\beta_{\cal P'}\left(\frac{\nu}{m}\right)^{\mu -1}+c_0+c_1\ln\left(\frac{\nu}{m}\right)+c_2\ln^2\left(\frac{\nu}{m}\right),\label{sig0pp}
\ea 
where $\nu$ is the laboratory energy of the incoming proton (anti-proton), $m$ the proton mass, and the `Regge intercept' $\mu=0.5$. The predictions for the $pp$ and $\bar pp$ total cross sections are shown in Fig. \ref{fig:pp}. The dominant $\ln^2(s)$ term in the total cross section (\eq{sig0pp}) saturates the Froissart bound \cite{froissart}; it controls the asymptotic behavior of the cross sections. BH made a  simultaneous fit\cite{BH} to the $pp$ and $\bar pp$ data for the $\rho$ value, the ratio of the real to the imaginary forward scattering amplitudes, shown in Fig. \ref{fig:rho}. From \eq{rhopmpp} and \eq{sig0pp}, we see that in the limit of $s\rightarrow \infty,\ \rho \rightarrow 0$ as $1/\ln s$, (albeit very slowly), a necessary condition for a black disk. Although the $\rho$-values are essentially the same for $\bar pp$ and $pp$ for $\sqrt s> 100$ GeV, at the highest accelerator energies, $\rho$ only changes from 0.135 at 7000 GeV to 0.132 at 14000 GeV.  Clearly, we are no where near asymptopia, where $\rho=0$. 

With two low energy constraints at 6 GeV and 4 parameters, precise values for $c_0$ and $\beta_{\cal P'}$ could be obtained\cite{BH}. The fitted values for the coefficients of $\sigma^0(\nu)$ of \eq{sig0pp} for the fit for $6\le \sqrt s\le 2000$ GeV are listed in Table \ref{tab:sigma0}.   Evaluating \eq{sig0pp} at 57000 GeV, we predict $\sigtot=134.8\pm 1.5$ mb for $pp$ interactions. We note that  $c_2$, the coefficient of $\ln^2(s)$, is well-determined, having a statistical accuracy of $\sim 2\%$. Thus, experimental data show that the Froissart bound is satisfied for  total cross sections $\sigtot$ for both $\bar p p$ and for $pp$ in the energy interval $6\le \sqrt s\le 2000$ GeV. 

\begin{figure}[h,t,b] 
\begin{center}
\mbox{\epsfig{file=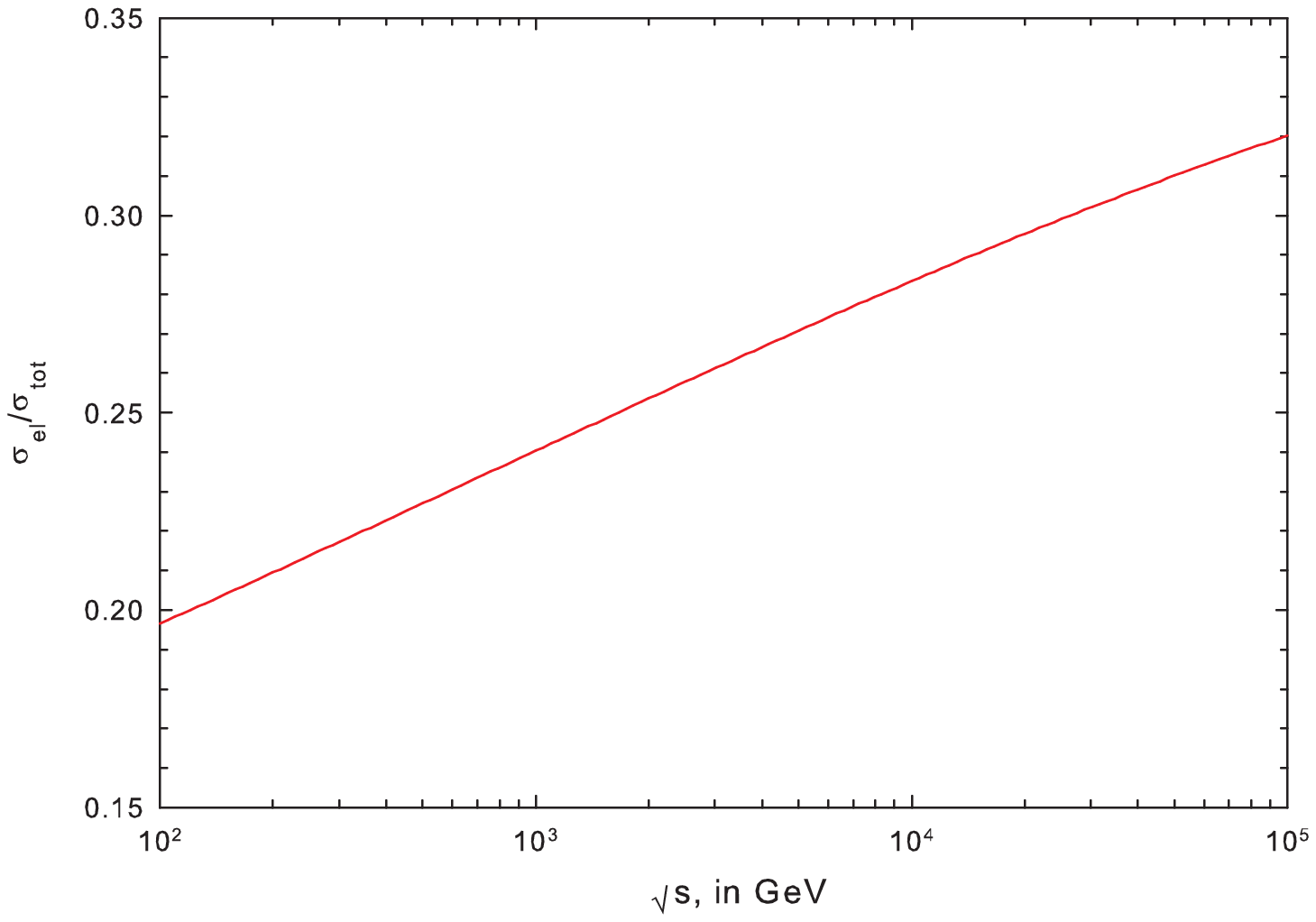
,width=3in%
,bbllx=60pt,bblly=255pt,bburx=500pt,bbury=560pt,clip=%
}}
\end{center}
\caption[]{
 The  $r$-value, the ratio of  $\sigel/\sigtot$,  vs. $\sqrt s$, the cms energy in GeV. \label{fig:r}
}
\end{figure}

\begin{table}[h,t]                   
%
\def\arraystretch{1.5}            

\begin{center}
\caption[]{Values of the parameters for the even amplitude, $\sigma^0(\nu)$, using 4 FESR analyticity constraints (taken from Ref. 
\cite{BH})
\label{tab:sigma0}
}
\vspace{.2in}
\begin{tabular}[b]{||l||l||l||l||}
\hline\hline
$c_0$=$37.32$ mb,
&$c_1$=$-1.440\pm 0.070$ mb,
&$c_2$=$0.2817\pm 0.0064$ mb,
&$\beta_{\cal P'}$.=$37.10$ mb\\
\hline\hline
\end{tabular}
\end{center}
\end{table}
\def\arraystretch{1}  
{\em Aspen Model:} The Aspen model \cite{aspenmodel} is an eikonal model that describes experimental $\bar pp$ and $pp$ data for $\sigtot$, $\rho$ and the slope parameter $B\equiv d[\ln d\sigel/dt]_{t=0}$, the logarithmic derivative of the forward  differential elastic scattering cross section, where $t$ is the square of the 4-momentum transfer. Among many other quantities, it allows one to accurately predict the ratio $r= \sigel(\nu)/\sigtot(\nu)$, i.e.,  the ratio of the elastic to total cross section for both $\bar pp$ and $pp$, as a function of energy, where again, the total cross sections have been anchored at 6 GeV by FESR constraints \cite {FESR}.  Details of the model are given in Ref. \cite{aspenmodel,physicsreports}.  As is the case of the total cross sections, the values for $r$ are essentially identical for $\bar pp$ and $pp$ for cms energies $\sqrt s \ge 100$ GeV. The ratio $r$ is plotted in Fig. \ref{fig:r}. Again, we see that we are far from asymptopia,  where the black disk model implies a ratio $r=1/2$, whereas at 57000 GeV, we predict $r\sim 0.32$.

{\em Inelastic cross section:} We are now ready to evaluate $\sigin(\nu)\equiv(1-r(\nu))\sigma^0(\nu)$ {\em numerically} for $\sqrt s\ge 100$ GeV, using $r(\nu)$ obtained above, together with  the fitted  even amplitude cross section $\sigma^0(\nu)$ of \eq{sig0pp} determined by the parameters of Table \ref{tab:sigma0}. Since the approach is at this point purely numerical, we decided to fit the inelastic numbers with the {\em same analytical parameterization} as was used for the total cross section $\sigma^0(\nu)$ in \eq{sig0pp}. The analytic expression for the  even amplitude high energy inelastic cross section $\sigma^0_{\rm inel}(\nu)$ given by
\ba
\sigma^0_{\rm inel}(\nu)&\equiv&\beta_{\cal P'}^{\rm inel}\left(\frac{\nu}{m}\right)^{\mu -1}+c_0^{\rm inel}+c_1^{\rm inel}\ln\left(\frac{\nu}{m}\right)+c_2^{\rm inel}\ln^2\left(\frac{\nu}{m}\right)\\
&=& 62.59\left(\frac{\nu}{m}\right)^{-0.5}+24.09+0.1604 \ln\left(\frac{\nu}{m}\right)+ 0.1433 \ln^2\left(\frac{\nu}{m}\right) \ {\rm mb}
\label{finalinelastic}
\ea 
accurately reproduces the numerical values of $\sigin(\nu)$ to better than 
4 parts in $10^4$ over the energy range $100\le\sqrt s \le 100000$ GeV. This new result for $\sigma^0_{\rm inel}(\nu)$ implies that the Froissart bound is also saturated for the  high energy  {\em inelastic}  cross sections in the energy interval $100\le \sqrt s\le 57000$ GeV, and is shown as the lowest curve in Fig. \ref{fig:pp}. This result was anticipated theoretically, using analyticity and unitarity,  by Andre Martin \cite{martin2}.

In Fig. \ref{fig:pp}. the lowest energy inelastic cross section datum  points, the (red) diamonds, are $\bar p p$ inelastic cross sections. The LHC 7000 GeV $pp$ inelastic cross section data points are the  (black) circle  from CMS \cite{CMS} and the (green) square  from Atlas \cite{Atlas}, slightly separated for visual purposes.  The (blue) diamond is the Auger inelastic cross section \cite{Auger} for a 25\% ${\rm He}^4$ contamination of their $\sigma_{\rm in}^{\rm p-air}$ cross section at 57000 GeV. We emphasized that {\em none} of these experimental inelastic cross sections were used in our fits that predicted high energy inelastic cross sections. At 7000 GeV our prediction is $\sigin = 69.0\pm1.3$ mb and at 57000 GeV $\sigin = 92.9\pm1.6$ mb.  Inspection of Fig. \ref{fig:pp} shows that the $\ln^2(s)$ fit of \eq{finalinelastic} for $\sigma^0_{\rm inel}(\nu)$ is in excellent agreement with all experimental data, up to the highest possible energy.

{\em Evidence for a black disk:} It is unlikely that there will ever be higher energy measurements for $\sigin$ for either $\pbar p$ or $pp$ collisions, yet our results show that present measurements are far from asymptopia. Nevertheless, the data give us a consistent picture of asymptopia by the compelling evidence that both the elastic and inelastic cross sections saturate the Froissart bound.  The addition of the inelastic cross section of \eq{finalinelastic}  going as $\ln^2 s$ now allows us to explore asymptopia {\em experimentally}; we find the limit of $\sigin(s)/\sigtot(s)$ as $s\rightarrow \infty$ simply  by  taking the ratio of the $\ln^2(s)$ terms in \eq{finalinelastic} and \eq{sig0pp}. We find the experimentally-determined value at infinity,
\ba
\stackrel{{\textstyle \lim}}{ s\rightarrow \infty} {\sigin(s)\over \sigtot(s)}&=& {c_2^{\rm inel}\over c_2}
= 0.509\pm 0.011,\label{diskratio}
\ea
a result compatible with the ratio 1/2 predicted for a black disk. Satisfying this ratio of the inelastic to the total cross section at infinity  gives us the first experimental evidence that the proton becomes an expanding black disk at asymptopia.  We have already shown that the second  condition,  $\rho =0$, i.e., the amplitude is imaginary,  is also satisfied. The model of Troshin \cite {Troshin} in which the elastic scattering dominates over the inelastic is thus falsified, whereas the models \cite{ChouYang,Bourrely} in which the proton becomes a black disk asymptotically are now justified experimentally. 
 
{\em Properties of a black disk}: In impact parameter space $b$, the elastic and total cross sections are given by 
\ba
\sigel&=&4\int d^2b\,\, |a(b,s)|^2,\qquad
\sigtot=4\int d^2b\,\, {\rm Im}\,a(b,s).
\ea
The amplitude $a(s,b)$ of the black disk of radius $R$ is given by
\ba
a(b,s)&=& {i\over 2},\quad 0\le b\le R, \qquad
a(b,s)=0, \ \quad b>R,
\ea
so that (for details, see Ref. \cite{BC})
\ba
\sigtot&=&2\pi R^2, \qquad
\sigin=\sigel= \pi R^2,\qquad
{\sigin\over \sigtot}=0.5,\qquad
{d\sigma_{\rm el}\over dt}=\pi R^4 \left[ {J_1(qR)\over qR}\right]^2,
\quad {\rm where\ } q^2=-t.
\ea
Using analyticity and unitarity,  Andre Martin  has recently found a more rigorous {\em inelastic} hadron-proton bound \cite{martin2}, using $t=(2m_\pi)^2$, i.e.,
\ba
\sigin < {\pi\over 4 m_\pi^2}\ln^2s,\quad {\rm so\ that\ }\sigtot<{\pi\over 2 m_\pi^2}\ln^2s \label{limit}
\ea
 where for the total cross section bound we have invoked the black disk ratio of 2 to 1. The use of $m_\pi$ in the two-particle mass $M=2m_\pi$ is clearly wrong experimentally, since ${\pi\over 2 m_\pi^2}\ln^2(\nu/m) = 31.23 \ln^2(\nu/m)  
$ mb, whereas experimentally we have obtained $c_2\ln^2(\nu/m)=0.2817 \ln^2(\nu/m)$ mb, a cross section two orders of magnitude smaller, implying that the scale is not set by the pion mass but by a mass scale one order of magnitude larger.
Reinterpreting  $M=2 m_\pi$   in \eq{limit} as the lowest-lying glueball mass which we call $M_{\rm glueball}$, we find $
M_{\rm glueball}=\left({2\pi/ c_2}\right)^{1/2}=2.97\pm 0.03\quad {\rm GeV}.$ Obviously, the definition of this scale is still arguable.

Also, if the asymptotic proton is a black disk of gluons, the high energy behavior is flavor blind and the coefficient of the $\ln^2 s$ term is the same for all reactions, from $\pi p$ to $\gamma p$ scattering. Support for this claim comes from both the COMPETE group\cite{COMPETE} and Ishida and Igi \cite{ishida}.
 
{\em Conclusions:} We find that the $\ln^2 s$ Froissart bound for the proton for $\sigtot$ \cite{froissart} and $\sigin$ \cite{martin2} is saturated and that at  infinite $s$, (1) the experimental ratio $\sigin/\sigtot= 0.509\pm0.011$, compatible with the black disk ratio of 0.5 and (2) the forward scattering amplitude is purely imaginary. We thus conclude that the proton becomes an expanding black disk at sufficiently ultra-high energies that are  probably never accessible to experiment. The theory for these bounds is predicated on the pillar stones of analyticity and unitarity, which have now been experimentally verified up to 57000 GeV.  Further, since $\sigtot$ has been extrapolated up from the Tevatron, we expect no new large cross section  physics between 2000 and 57000 GeV.

Finally, the lowest-lying glueball mass is measured to be $M_{\rm glueball}=2.97\pm 0.03\quad {\rm GeV}$. Reproducing these experimental results will be a task of lattice QCD.

{\em Acknowledgments:} In part,  F. H. is supported by the National Science Foundation  Grant No. OPP-0236449, by the DOE  grant DE-FG02-95ER40896 and  by the University of Wisconsin Alumni Research Foundation. M. M. B. thanks  the Aspen Center for Physics, supported in part by NSF Grant No. 1066293,  for its hospitality during this work.

\end{document}